\documentclass[twocolumn,pra,showpacs]{revtex4}
\usepackage{amssymb,amsmath,amsfonts,bm}
\usepackage{graphicx}
\usepackage{color}

\begin{document}

\date{\today}

\title{Oscillations in turbulence-condensate system}

\author{Pearson Miller$^1$, Natalia Vladimirova$^2$,
and Gregory Falkovich$^3$}
\affiliation{$^1$Yale University, Department of Physics, New Haven CT 06511\\
\mbox{$^2$University of New Mexico, Department of Mathematics and Statistics, Albuquerque NM 87131}
$^3$Weizmann Institute of Science, Rehovot 76100 Israel}

\begin{abstract}
We consider developed turbulence in the Gross-Pitaevsky model where
condensate appears due to an inverse cascade.  Despite being fully
turbulent, the system demonstrates non-decaying periodic oscillations
around a steady state, when turbulence and condensate periodically
exchange a small fraction of waves.  We show that these collective
oscillations are not of a predator-prey type, as was suggested
earlier; they are due to phase coherence and anomalous correlations
imposed by the condensate.
\end{abstract}


\maketitle


Understanding the interaction of turbulence and a coherent flow is an
important problem in turbulence studies in fluid mechanics and beyond,
both from fundamental and practical perspectives. In fluids, coherent
flows are system-size vortices or zonal flows of different
profiles~\cite{Shats,CKL}, which are known to diminish the turbulence
level, change its nature, and make its statistics more
non-Gaussian~\cite{Shats,NP}. Here we consider arguably the simplest
case of a turbulent system with a condensate: the coherent part is
expected to be a constant  and turbulence to consist of weakly
interacting waves.

The Nonlinear Schr\"odinger equation (NSE), also known as the
Gross-Pitaevsky equation, provides a universal description of the
evolution of any nonlinear, spectrally narrow wave packet \cite{Fal,ZLF,Naz}:
\begin{equation}
      i\psi_t + \nabla^2 \psi - |\psi|^2\psi = 0.
      \label{NSE}
\end{equation}
As such, it can provide a model of nonlinear behavior in a wide range
of physical systems, from locally interacting bosons to plasmas and
fluids. The addition of damping and driving causes the NSE to
exhibit turbulence, and its universality makes it a major subject of
interest in the study of wave turbulence
\cite{ZLF,Naz,Opt,DF,no}. Because the equation conserves both total
energy and wave action,
\[
  {\cal H} =  \textstyle{\int}\left(|\nabla\psi|^2 +
         \textstyle{\frac{1}{2}}|\psi|^4\right)\,d {\bf r}, \qquad
   {\cal N} =  \textstyle{\int}|\psi|^2\,d{\bf r},
\]
it has the potential to form both a direct and inverse turbulent
cascade.  In a system with external forcing, the inverse cascade leads
to the growth of a single coherent mode known as the spectral
condensate, which comes to dominate the dynamics of the system.
Understanding the interaction between a large condensate and other
spatial modes of a system governed by the NSE is crucial to a deeper
understanding of wave turbulence in such a system.

Numerical simulations of wave turbulence in the NSE suggest the
existence of 
collective oscillations of the turbulence-condensate
system~\cite{Opt,VDF}.  During these oscillations, a small fraction of
wave action is periodically converted from the condensate to the
turbulent part of the spectrum, with total wave action unchanged.
Such oscillations were predicted to take place when broad turbulent
spectra coexist with a sharp spectral peak~\cite{Fal84}. For
three-wave interaction, a simple model of a predator-prey type that
describes the evolution of the total numbers of waves in the two
groups has the form: $dN_0/dt=-b N_0+N_0n$, $dn/dt=\gamma n - N_0n$,
which gives the oscillations with the frequency $\sqrt{\gamma
  b}=\sqrt{\bar n\bar{N_0}}$~\cite{Fal84}.  For the oscillations in
NSE with a condensate a similar model was suggested~\cite{Opt}:
$dN_0/dt=-b N_0 + N_0 n^2$, $dn/dt=\gamma n - N_0^2 n$. The most
evident defect of this model is that it does not conserve the total
number of waves. Moreover, its predictions for the steady-state values
and the frequency of oscillations are in disagreement with the data,
see~\cite{VDF1}.  It was also pointed out in~\cite{VDF1} that the
account of phase coherence (or anomalous correlations) is needed to
describe the oscillations (for the account of anomalous correlations,
see also~\cite{W}).  Unfortunately, the model with anomalous
correlations was not treated properly in~\cite{VDF1}.  Here we derive
the simplest model of this type, based on anomalous correlations in
the three-wave system of two counter-propagating waves interacting
with the condensate.  We study the model analytically and compare the
results with the oscillations of individual modes extracted from the
direct numerical simulations of the turbulence in the NSE.  We
demonstrate that the three-wave model provides reasonable explanation
for collective oscillations.


\begin{figure*}
\includegraphics[width=0.75\textwidth]{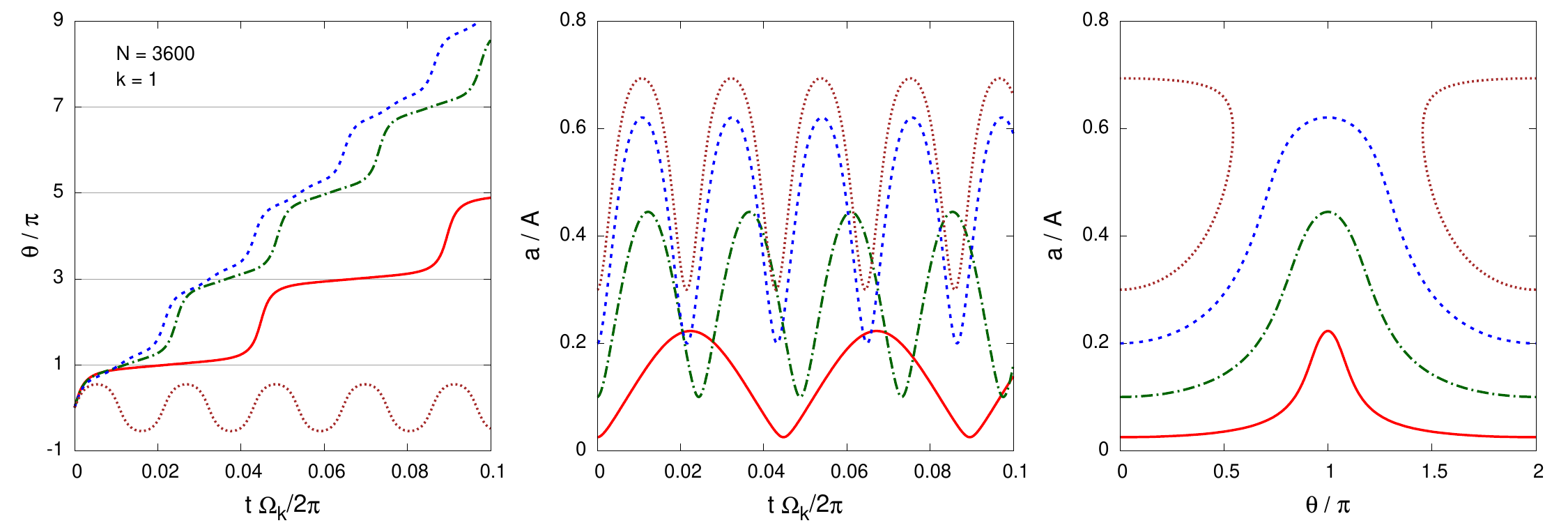}
\caption{(Color online) Numerical solution of ODE system
  (\ref{ODE_n_full})-(\ref{ODE_th_full}) with $N=3600$ and $k=1$:
  phase difference (left), the normalized amplitude $a/A =
  \sqrt{n/N}$ (middle), and phase portrait (right).  Different curves
  correspond to different initial conditions.}
\label{fig:ODE_full}
\end{figure*}

To start, we show that an important element of turbulence against the
background of a condensate must be anomalous correlations, i.e. phase
coherence between waves running in opposite directions.
Let us denote $N=\overline{|\psi|^2}$ and $N_0=|\overline{\psi}|^2$.
The simplest condensate is a spatially-uniform field, $\Psi=\sqrt{N_0}\exp(-iN_0t)$,
which is an exact solution of (\ref{NSE}). Small over-condensate
fluctuations satisfy
\begin{equation}i\dot\psi_k=(k^2+2N_0)\psi_k + \Psi^2\psi_{-k}^*.
\label{linear}
\end{equation}
This equation gives  the Bogolyubov dispersion relation,
\begin{equation}
  \Omega_k^2=2N_0k^2+k^4 \ ,
  \label{Bog}
\end{equation} for a pair of counter-propagating waves
$\psi_{\pm k} \propto \exp(-iN_0t \mp i\Omega_kt)$.

To look into the effective dynamics of both the number of waves and the
condensate, let us assume for simplicity that the condensate interacts
with only two contra-propagating waves with amplitudes
\[
        \psi_{\pm k} = \sqrt{n} \exp(-iN_0 t  \pm ikx + i\phi_{\pm k})
\]
and phase difference  $\theta = 2\phi_0 - \phi_k - \phi_{-k}$.
\begin{equation}
      H = 2 k^2 n + \textstyle{\frac{1}{2}} N^2 + 2n(N-2n)(1 + cos \theta) + n^2.
      \label{Ham}
\end{equation}
That gives the equations \cite{Fal}:
\begin{eqnarray}
     \dot{n} &=& 2n (N - 2n) \sin \theta, \label{ODE_n_full}\\
     \dot{\theta} &=& 2k^2 + 2(N-4n)(1+\cos\theta) + 2n. \label{ODE_th_full}
\end{eqnarray}

Combining (\ref{Ham}) and (\ref{ODE_n_full}) one can obtain the
equation for wave amplitude, $\dot{n} = \sqrt{f(n)}$, where $f(n) =
7n^4 + (12k^2 - 4N)n^3 + (3N^2 -6H -4k^2 - 8Nk^2)n^2 + (4Hk^2 + 4HN - 2k^2N^2)n +
(HN^2 - H^2 - \frac{N^2}{4})$.  The solution of this equation can be expressed via the elliptic
functions: $n(t) = n_0 + \frac{1}{4}f'(n_0) \left( {\cal
  P}(t, g_2, g_3) - \frac{1}{24}f''(n_0) \right)^{-1}$.  Here $n_0 =
-k^2 + \sqrt{H + k^4 - \frac{N^2}{4}}$ is the solution of $f(n) = 0$, $\cal P$ is the
Weierstrass elliptic function, and $g_2$ and $g_3$ are two invariants
of $\cal P$ determined by coefficients in $f(n)$. 

The typical solutions of system (\ref{ODE_n_full})-(\ref{ODE_th_full})
are shown in Fig.~\ref{fig:ODE_full} for a wide range of initial
conditions.  Motivated by application with very large condensate, we
are mostly interested in limit of $n/N \ll 1$ (solid line in
Fig.~\ref{fig:ODE_full}).  This limit is characterized by longer
periods of oscillations, cusped shape of $a(t)$ curves (where $a
\equiv \sqrt{n}$), and by open trajectories in the phase space.  The
system spends most of its time around $\theta=\pi$ state, avoiding
both stable points: $\theta = 0$, $n=(4N + k^2)/14$ with its
unrealistically high $n/N$ ratio, and the unphysical $\theta = \pi$,
$n = -\frac{1}{2}k^2$.

In the limit of $n \ll N$, the system
(\ref{ODE_n_full})-(\ref{ODE_th_full}) reduces to $\dot{n} = 2nN \sin
\theta$, $\dot{\theta} = 2k^2 + 2N(1+\cos\theta)$ resulting in
\begin{eqnarray}
   n(t) &=& n(0) \left( 1 + \frac{2N}{k^2} \sin^2 \Omega_k t \right),
        \label{n(t)}\\
   \theta(t)  &=& 2 \arctan \left( \frac{\Omega_k}{k^2} \tan \Omega_k t  \right),
        \label{th(t)}\\
   n(\theta) &=&  n(0) \frac{2N + k^2}{ N (1 + \cos\theta) + k^2}.
        \label{n(th)}
 \end{eqnarray}
Here, $\Omega_k = \sqrt{2Nk^2 + k^4}$ is the Bogolubov frequency, and
$n(0) = n|_{\theta=0}$ is the constant of integration.  The second
constant of integration shifts the solution in time; it is selected so
that $\theta|_{t=0}=0$.  The solution (\ref{n(t)})-(\ref{n(th)}) has
the frequency of oscillations $2\Omega_k$, the stair-like time dependence of
phase, and the cusped shape of amplitude, $a=\sqrt{n}\propto |\sin
\Omega_k t|$, as suggested by the numerical
solution of (\ref{ODE_n_full})-(\ref{ODE_th_full}) shown in
Fig.~\ref{fig:ODE_full}.  Next, we demonstrate that $n \ll N$
approximation is well justified for the levels of condensate typical
for NSE turbulence.


\begin{figure}[b]
 \includegraphics[width=0.48\textwidth]{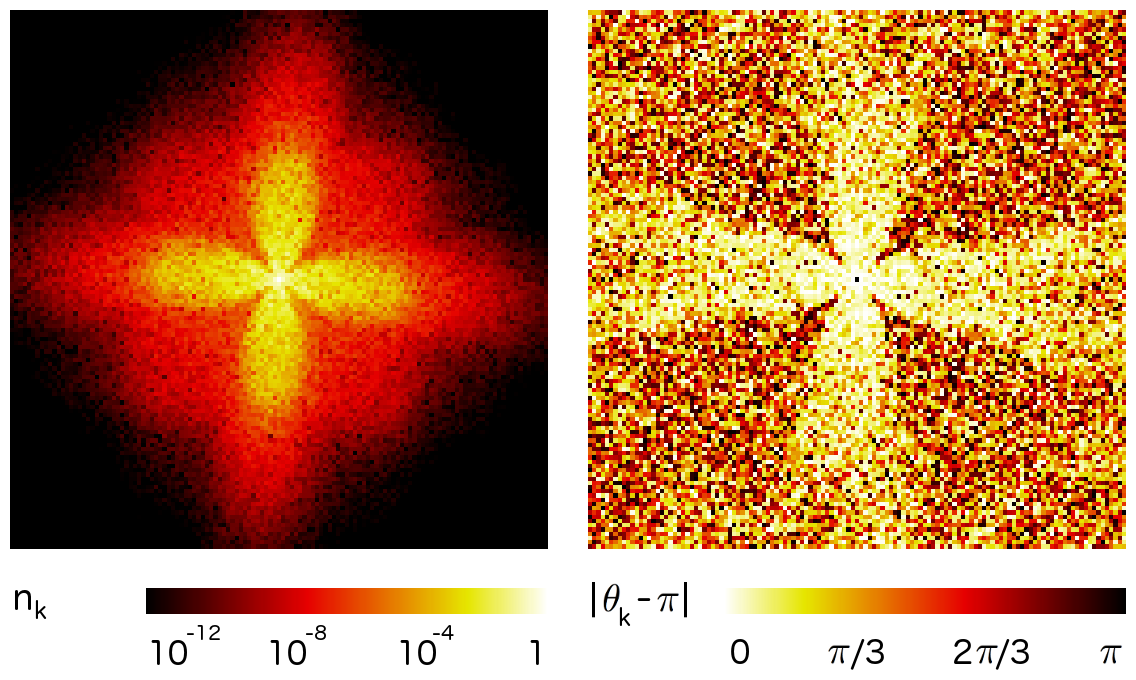}
 \caption{(Color online) Spectrum of NSE turbulence $n_k$ (left) and
   phase difference $\theta_k$ (right), $N=3600$.  }
  \label{fig:spectrum}
\end{figure}

\begin{figure*}
\includegraphics[width=0.75\textwidth]{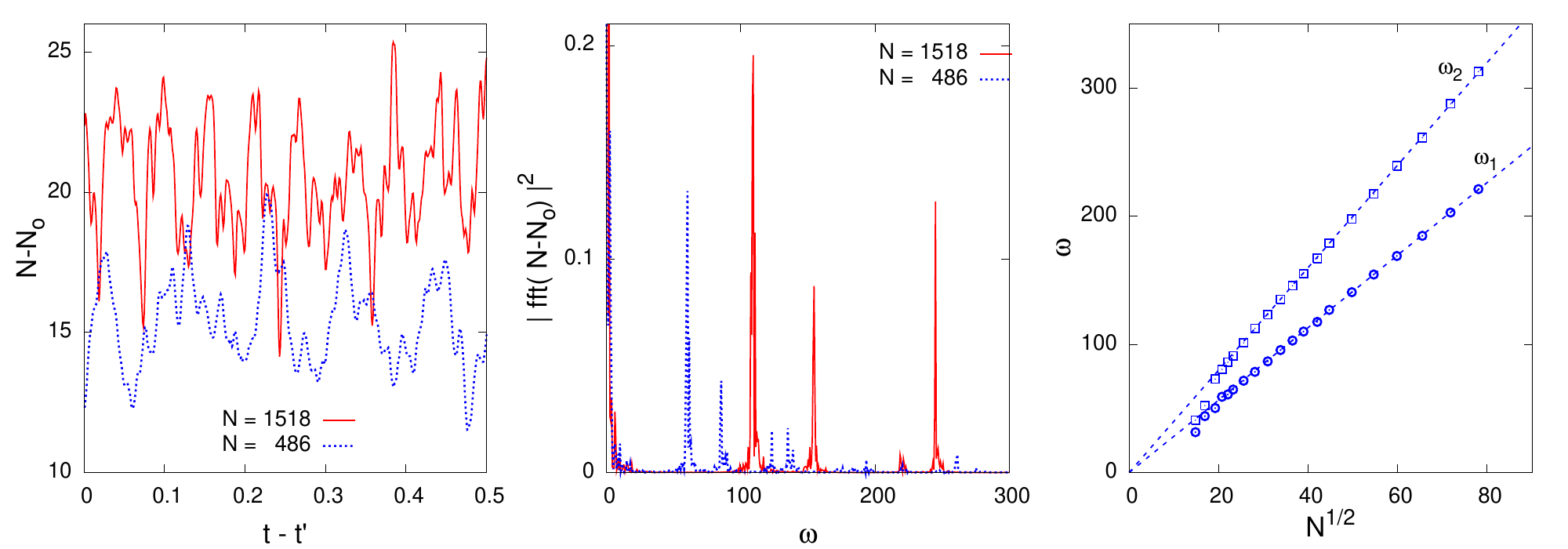}
\caption{(Color online)
Left: oscillations in over-condensate zoomed
around times $t'=800$ ($N=486$) and $t'= 3500$ ($N=1518$).  Middle:
frequency spectrum of the time-dependence of the number of
over-condensate waves for the same run in time intervals $[800,810]$
and $[3500,3510]$.  Right: first two frequencies as function of the
number of waves next to lines with slopes 4 and $2\sqrt{2}$.  }
\label{fig:oscil}
\end{figure*}

We now compare the solution of the ODE
model~(\ref{ODE_n_full})-(\ref{ODE_th_full}) with the results of
numerical simulations of developed turbulence in nonlinear Schr\"odinger
equation~(\ref{NSE}).  Our numerical simulations, based on the 4th
order fully dealiased split-step method~\cite{agrawal,yoshida}, are
set up similarly to~\cite{DF} and are described in detail
in~\cite{VDF}.  Simulation are done in periodic $2\pi \times 2 \pi$
domain.  Extra terms are added to Eq.~(\ref{NSE}) to model forcing ---
the large scale multiplicative pumping and small scale damping.  In
simulations where thermal equilibrium is used as initial conditions,
the pumping is needed to gradually raise the wave action and develop
the condensate.

In such simulations, as the wave action increases with time, the
system undergoes the series of phase transitions~\cite{VDF}.
The ``phases'' are distinguished by different symmetries of the
spectrum and by different spatial patterns in over-condensate
fluctuations observed on small scales.  The shape of the spectrum
changes from radially symmetric to two-petal spectrum at $N \sim 500$,
to three-petal spectrum at $N \sim 1000$, and to four-petal spectrum
at $N \sim 2000-4000$.  Further transitions are possible.  The
transitions are related to the decrease of the angle of wave
interactions, which causes the spectrum to break into a large number
of narrow-angle bands.

The typical four-petal spectrum, $n_k$, is shown in
Fig.~\ref{fig:spectrum} together with phase differences, $\theta_k$.
Even though the phases of different modes appear to be random, the
modes with opposite wavenumbers are pairwise correlated by their
interactions with the condensate.

\begin{figure*}
\includegraphics[width=\textwidth]{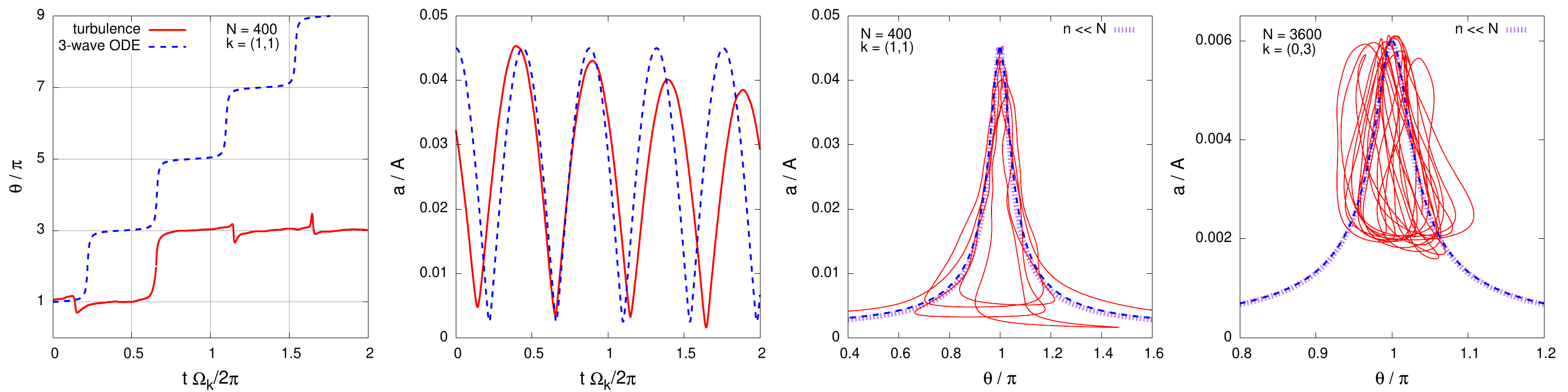}
\caption{(Color online)
  First three panels: comparison of oscillations of ${\bf k} =
  (1,1)$ mode extracted from simulation of turbulence at $N=400$ with
  the prediction of three-wave model; from left to right, phase,
  amplitude, the phase portrait are shown. Fourth panel:
  comparison of phase portraits for $N=3600$ and ${\bf k}=(0,3)$.
  Here, as in Fig.~\ref{fig:ODE_full}, $a/A = \sqrt{n/N}$.
}
\label{fig:amp_phase}
\end{figure*}

Condensate oscillations at the frequencies of individual modes add up
to the overall fluctuations around condensate's average, shown
in~Fig.~\ref{fig:oscil}.  The highest contribution to the signal is
from the modes with lowest wave numbers.  Our three-wave model
predicts that the modes oscillate at twice the Bogolyubov frequency.
Indeed, the NSE simulations show that the dominant frequency of
condensate-turbulence oscillations, $2\sqrt{2 N_0}$, is the
frequency of $|{\bf k}|=1$ modes, while the frequency of the second
harmonic, $4\sqrt{N_0}$, correspond to $|{\bf k}|^2=2$ modes.  These
frequencies (which are much smaller than the frequency of the phase
rotation of the condensate $N$) essentially do not depend on the level
of over-condensate fluctuations.  They are clearly seen in the
oscillations of the condensate amplitude and of the normal correlation
functions.

Our three-wave model describes not only the frequencies but also the
time dependence of the oscillations.  The time dependence of the phase
and the amplitude of an individual mode extracted from a NSE
simulation is compared to the prediction of the model
(\ref{ODE_n_full})-(\ref{ODE_th_full}) in Fig.~\ref{fig:amp_phase}.
Here, the $n/N$ ratio is small, and the reduced model
(\ref{n(t)})-(\ref{n(th)}) works as good as
the full one, as illustrated in phase portrait section of
Fig.~\ref{fig:amp_phase}.  As predicted by the model, the amplitude
$a_k$ has cusped shape, while the phase is localized around $\theta =
\pi$.

The biggest difference between the NSE modes and the model is the
phase portrait.  The model predicts monotonous increase of the phase
from $(2j-1)\pi$ to $(2j+1)\pi$, while the phase of the NSE modes
oscillates around $\pm\pi$ in closed loops.  Apparently, the
interaction with other modes, unaccounted in our simple model, leads
to phase locking.

\begin{figure}
 \includegraphics[width=70mm]{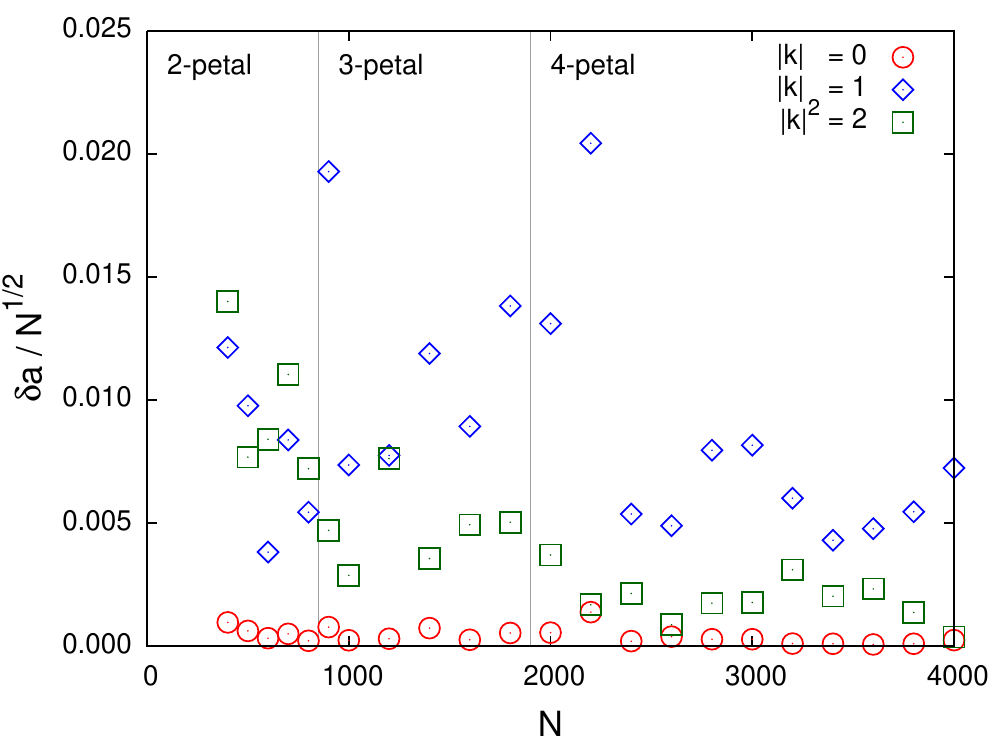}
 \caption{(Color online)
   Amplitude of oscillations of low modes for different levels
   of condensate is averaged over time interval $[10,12]$
   in simulations with preset condensate. 
   }
  \label{fig:manyN}
\end{figure}

The amplitude in the three-wave model is determined up to the constant
of integration. In other word, the level of fluctuations must be
obtained from the turbulence data.  Our study of NSE turbulence at
different levels of condensate indicates that the amplitude of lower
modes remain roughly constant for the wide range of $N$.  This also
means that $n/N$ decreases as the condensate level increases, as shown
in Fig.~\ref{fig:manyN}. (There is a possibility that $n/N$ is
somewhat larger near phase transition but this topic needs additional
investigation.)  In all cases considered, the ratio $n/N$ is small,
well within the applicability limits of the reduced model.  At higher
$N$, the loops in the phase portrait are reconnecting tighter and
closer to $\theta=\pi$.  It is also interesting that amplitudes of
non-condensate modes are larger than fluctuations of the condensate.

Regarding the data presented in Fig.~\ref{fig:manyN}, we shall make
the following technical comment, unaddressed in~\cite{VDF}.  The
states, similar to the state shown in Fig.~\ref{fig:spectrum}, can be
achieved in numerical simulation by using initial conditions where the
condensate is superimposed on top of the thermal equilibrium
spectrum.  For such systems, it take only a few linear time units to
relax to quasi-equilibrium with the appropriate symmetry, in contrast
to thousands of time units of evolution needed for systems without
preset condensate.  Moreover, as shown in Fig.~\ref{fig:oscil}, the
condensate-turbulence oscillations are very fast and can be studied in
simulations without forcing (except for the small damping for smooth
transition to de-aliased region).

To summarize, at the large level of condensate the three-wave model
capture the following features of the turbulence-condensate
oscillations: (i) the frequency of oscillations is twice the Bogolubov
frequency, (ii) the system spends most of its time around $\theta=\pi$
state, and (iii) the amplitude as a function of time has a non-trivial,
cusped shape.  To describe the shape of the phase oscillations around
$\pm \pi$, additional mechanisms need to be included.

This research was supported by the Kupcinet-Getz International Science
School and by the grants of the BSF, ISF and the Minerva Foundation
funded by the German Ministry for education and research.  
Work of N.V. was supported by NSF grants PHY 1004118 and PHY 1004110.


\end{document}